\documentstyle[epsf,prl,aps,multicol]{revtex}

 \input{psfig}

\begin{document}
%\begin{multicols}
\draft

%\tighten
\title{Electron localization in sound absorption
oscillations in the quantum Hall effect regime}

\author{
I. L. Drichko, A. M. Diakonov, A. M. Kreshchuk, T. A. Polyanskaya,
I. G. Savel'ev, I. Yu. Smirnov, and A. V. Suslov}
\address{A. F. Ioffe Physico-Technical Institute of Russian
Academy of Sciences, Polytechnicheskaya 26, 194021, St.Petersburg,
Russia}

\date{\today}
\maketitle

\begin{abstract}
The absorption coefficient for surface acoustic waves in a
piezoelectric insulator in contact with a GaAs/AlGaAs
heterostructure (with two-dimensional electron mobility $\mu=
1.3\times 10^5 cm^2/V\cdot s)$ at $T$=4.2K) via a small gap has
been investigated experimentally as a function of the frequency of
the wave, the width of the vacuum gap, the magnetic field, and the
temperature. The magnetic field and frequency dependencies of the
high-frequency conductivity (in the region 30-210 MHz) are
calculated and analyzed. The experimental results can be explained
if it assumed that there exists a fluctuation potential in which
current carrier localization occurs. The absorption of the surface
acoustic waves in an interaction with two-dimensional electrons
localized in the energy "tails" of Landau levels is discussed.
\end{abstract}
%\bigskip

%\narrowtext

\begin{multicols}{2}

\section{INTRODUCTION}

The problems of quantum interference of electrons, i.e., weak
localization in a two-dimensional electron gas (2DEG), just as
problems of strong charge-carrier localization (associated with
defects of the crystal lattice), occupy a central place in
two-dimensional nanoelectronics, but they have still not been
adequately investigated. The characteristic features of
localization in a 2DEG can be studied by investigating rf
conductivity of a 2DEG in the quantum Hall effect regime \cite{1}

One proven method for investigating rf conductivity is the
acoustic method, which makes it possible to measure the rf
conductivity of semiconductors without any electrical contacts on
the sample. The crux of this method is that a piezoelectrically
active sound wave accompanied by an electric field undergoes
absorption as a result of interaction with electrons as it
propagates in the semiconductor, and this absorption is directly
related to the electrical conductivity of the sample.

Acoustic methods have been successfully used to study the rf
conductivity of a three-dimensional electron gas in doped
compensated semiconductors at low temperatures \cite{2,3}. It has
been shown that if the electrons are in a free (delocalized)
state, then the ultrasonic absorption coefficient of the electrons
in the semiconductor, which is a piezoelectric material, in a
magnetic field $H$ is determined completely by its dc conductivity
$\sigma_{xx}^{dc}(H)$. If, however, electrons are localized at
separate impurity centers or in a random fluctuation potential
produced by the impurities, as in the case of a strongly doped and
strongly compensated semiconductor, then the conduction mechanisms
in a constant electric field and in the rf field accompanying the
acoustic wave are different. In the static case the conductivity
$\sigma_{xx}^{dc}(H)$ is of a hopping character and the rf
conductivity $\sigma_{xx}^{ac}(H)$ can remain "metallic", and in
addition $\sigma_{xx}^{dc}(H)\neq\sigma_{xx}^{ac}(H)$. By studying
the ultrasonic absorption coefficient of the electrons as a
function of the magnetic field intensity, temperature, ultrasonic
frequency, and ultrasonic wave intensity in the three-dimensional
case it was possible to determine not only the character of the
carrier localization but also the corresponding numerical
parameters.

On this basis, it is helpful to extend the acoustic method of
investigating rf conductivity to semiconductor structures with a
2DEG. The interaction of surface acoustic waves (SAWs) with
two-dimensional electrons was first observed in GaAs/AlGaAs type
heterostructures \cite{4,5}. It was shown that the SAW energy is
absorbed as a result of this interaction, and the magnitude of the
absorption is determined by the conductivity of the 2DEG. For this
reason, in a quantizing magnetic field, when Shubnikov-de Haas
(SdH) oscillations appear, the absorption coefficient for the
surface acoustic wave also oscillates. The oscillations of this
coefficient are sharpest in the region of magnetic fields where
the quantum Hall effect regime obtains. In this connection, by
analogy with the three-dimensional semiconductors, the
characteristic features of localization in a 2DEG in a quantum
Hall effect regime can be studied by studying the temperature,
frequency and magnetic field dependencies of the sound absorption
and, which is especially important, this can be done by studying
the conductivity measured by a contact-free method.

Two experimental configurations are now employed. In the first one
SAWs are excited in the standard manner on the surface of a
piezoelectric crystal (for example, $LiNbO_3$), and the
experimental structure with the 2DEG is placed above the surface
of a piezodielectric crystal with a gap, whose width is less than
the wavelength. In the second configuration the experimental
structure itself, which must be a piezoelectric, serves as the
sound duct. Surface acoustic waves are excited in it and at the
same time the interaction of sound with the 2DEG is investigated.
The first configuration has the advantage that the gap eliminates
the deformation interaction of the two-dimensional (2D) electrons
with SAWs, and only the interaction with the electric fields
produced by the SAWs in the piezoelectric crystal remains.
However, this method is inconvenient because it is almost
impossible to determine the width of the gap by direct
measurements. In experiments employing the second configuration
there is no gap at all, but in this case the deformation
interaction of the 2D electrons with the SAWs cannot be ignored
\cite{6}.

The first configuration was used in \cite{4} and the second was
used in \cite{5,6,7,8}. In all of these investigations it was
concluded that the character of the dependence of the SAW
absorption coefficient of the two-dimensional electrons on the
magnetic field $H$ is determined by the dc conductivity
$\sigma_{xx}^{dc}(H)$ and its dependence on $H$. However, in
\cite{9}, where the SAW absorption of a 2DEG in a InGaAs/InP
heterostructure was measured in the first experimental
configuration with $T$=4.2 K, it was shown that this assertion is
valid only for high Landau-level numbers and in stronger magnetic
fields (i.e., small filling numbers $\nu=n hc/2eH$, where n is the
Hall electron density) the variation of the SAW absorption
coefficient as a function of $H$ is not determined by
$\sigma_{xx}^{dc}(H)$.

In the present work the dependencies of the SAW absorption
coefficient $\Gamma$ of 2D electrons on the magnitude and
direction (longitudinal and transverse with respect to the normal
to the plane of the 2DEG) of the magnetic field up to $H$=60kOe,
the SAW frequency in the range $f$=30-210MHz, and the temperature
in the range $T$=1.4-4.2K are investigated using the first
experimental configuration mentioned above for the purpose of
studying the rf conductivity and the character of electron
localization in structures of this type in the integral quantum
Hall effect regime. To get a better physical picture of the
phenomena, galvanomagnetic measurements of the components of the
dc resistivity tensor were performed as a function of the magnetic
field up to $H$=50kOe.

\section{EXPERIMENTAL PROCEDURE AND RESULTS}

The GaAs/AlGaAs structures investigated were prepared by
molecular-beam epitaxy. A layer of gallium arsenide with residual
impurity density $10^{15} cm^{-3}$ and thickness of the order of
$1 \mu m$ (the so-called buffer layer) was grown on a
semiinsulating GaAs (001) substrate of thickness $d=300 \mu m$.
Next, a 50-$\AA$-thick undoped layer of the solid solution AlGaAs
(spacer) and an approximately 500-$\AA$-thick layer of a solid
solution of the same composition but doped with silicon with an
electron density $n=10^{18} cm^{-3}$ were grown. The top (contact)
layer of the structure consisted of a 200-$\AA$ gallium arsenide
film doped with silicon [$n\simeq(3-4)\times 10^{18} cm^{-3}$.

The samples for the dc galvanomagnetic measurements and for
measurements by acoustic methods were prepared from neighboring
regions of the heterostructure. The measurements were performed
with dc current $I=10—20 \muÀ$ (which ensured that there would be
no heating effects) in the temperature range 1.8-4.2Ê on mesa
structures etched out in the form of a double cross with contact
pads. We present below the results of our investigations for a
structure with Hall density $n=1.6\times 10^{11} cm^{-2}$ and Hall
mobility $\mu = 1.3\times 10^5 cm^2/(V\cdots)$ at $T$=4.2K. The
density calculated according to the SdH oscillations was
$n_{SH}=6.8\times 10^{11} cm ^{-2}$ The results of the
measurements of the components $\rho_{xx}$ and $\rho_{xy}$ of the
resistivity tensor as a function of the magnetic field $H$ (in the
QHE region) at $T$=1.5K is presented in Fig.1.

A piezoelectrically active surface wave, excited on the Y-cut
surface of a $LiNbO_3$ crystal (see inset in Fig. 2) and
propagating in the $Z$ direction, was used for the acoustic
measurements. The SAW was produced and detected with interdigital
transducers with fundamental frequency 30, 90, and 150 MHz. In
addition, higher-order harmonics were also used in the experiment.
The sample was placed directly on the lithium niobate surface,
along which a SAW propagated, and detected with the aid of a
spring. The sample was located in a cryoduct in a vacuum chamber.
The vacuum gap between the piezodielectric and the sample,
determined by the roughness of their surfaces, is designated by
the letter $a$ in the diagram shown in the inset in Fig. 2. A
longitudinal electric field (along the direction of propagation of
the acoustic wave), which arises with the motion of the SAW in the
piezodielectric crystal, penetrates into the experimental object,
and the 2DEG is then located in an alternating electric field
whose frequency equals the frequency of the surface acoustic wave.
The interaction of the 2D electrons in the experimental
heterostructure with the SAW electric field causes the wave to
decay, and this decay is recorded by the transducer employed. The
independence of the absorption coefficient $\Gamma$ from the
intensity of the sound wave was checked in the experiment, i.e.,
the measurements were performed in a regime linear in the electric
field.

Figure 2 shows the dependence of the absorption coefficient for a
SAW with frequency $f$=30 MHz on the intensity of a magnetic
field, oriented in a direction normal to the surface of the
heterostructure, at $T$=4.2K and for different vacuum gap widths
$a$. Similar curves were also obtained for other frequencies and
temperatures. An interesting feature is observed on the curves in
Fig. 2: In strong magnetic fields the maxima of the absorption
$\Gamma$ split and the lower the temperature, the lower the
magnetic field intensity at which splitting is observed. The
maxima of $\Gamma$ are equally spaced as function of $1/H$ and for
large Landau numbers they correspond, with respect to the magnetic
field, to the minima of $\sigma_{xx}^{dc}(H)$; for higher fields
$H$, when $\sigma_{xx}^{dc}$ becomes small at the minima of the
oscillations ($<10^{-7} \Omega^{-1}$) the absorption maxima split
in two and the conductivity minimum (and also the center of the
Hall plateau) corresponds to an absorption minimum.

We also measured the absorption of SAWs in a magnetic field
oriented parallel to the surface of the sample. In this geometry
the SAW absorption oscillations were not observed. This confirms
the two-dimensional character of the absorption oscillations which
we observed in a transverse magnetic field.

An interesting feature of the absorption of SAWs in strong
magnetic fields is the inequality, observed for small filling
numbers, in the amplitudes of the SAW absorption maxima in the
presence of splitting. In \cite{5} this fact was attributed to the
inhomogeneity of the experimental samples. We performed a very
large number of measurements on several GaAs/AlGaAs differently
positioned (with different vacuum gaps) samples, and we observed
both a different asymmetry of the peaks and the same height of the
peaks on the same sample; this is illustrated in Fig. 2. It is not
yet clear what causes this effect.

\section{ANALYSIS OF THE EXPERIMENTAL DATA}

The equal spacing of the peaks in the dependence of the absorption
of surface acoustic waves (SAWs) on the quantity $1/H$ makes it
possible to determine the density of 2DEG from the period of the
oscillations by the standard method. The density obtained in this
manner equals, within the limits of accuracy of the experiment,
the density $n$ calculated according to the SdH oscillations.

The absorption coefficient $\Gamma$, which is associated with the
interaction of a SAW with two-dimensional electrons, in the
experimental configuration employed in our work was calculated in
Ref. 10, taking into account the diffusion of current carriers. If
the absorption coefficient $\Gamma$ is introduced as
$A=A_0exp(-\Gamma x)$, where $A_0$ and $A$ are the signal
amplitudes at the entrance and exit, respectively, and $x$ is the
length of the sample, then $\Gamma$ is (in $cm^{-1}$)

\begin{eqnarray}
\Gamma=\chi^{(1)} k \frac {\pi\sigma f_2(k)/\varepsilon_s v}
{1+[Dk/v+2\pi\sigma f_1(k)/\varepsilon_s v]^2}. \label{Eq1}
\end{eqnarray}
In \cite{10} $\chi^(1)\simeq K^2$ is the electromechanical
coupling constant of $LiNb0_3$; $k$ and $v$ are, respectively, the
SAW wave vector and velocity; $D$ is the diffusion coefficient;
$\varepsilon_s$ is the permittivity of the semiconductor; and
$\sigma$ is the conductivity of the 2DEG. We shall estimate the
contribution of diffusion to the absorption coefficient without
expanding the functions $f_1$ and $f_2$ (expressions for which are
presented in \cite{10}). For a degenerate 2DEG the diffusion
coefficient is

\begin{eqnarray}
D=\sigma \pi\hbar^2/e^2m^*=\pi\sigma a_B/\varepsilon_s, \label{Eq2}
\end{eqnarray}

where $m^*$ is the electron effective mass, and $a_B$ is the
effective Bohr radius (for electrons in GaAs $a_B \simeq 100
\AA$). We note that for all frequencies and vacuum gap widths
employed in the experiment, the value of $f_1$ calculated
according to Eq. (10) from \cite{10} ranged from 0.6 to 1.4, so
that the two terms in the denominator of (\ref{Eq1}) satisfy the
relation

\begin{eqnarray}
D \frac{k}{v}\ll\frac{2\pi\sigma}{v\varepsilon_s}f_1
 \label{Eq3}
\end{eqnarray}

(which reduces to the inequality $k\ll 2f_1 / a_B$), and it is
natural to use below the formula for the absorption coefficient
neglecting the diffusion term. If expression (\ref{Eq1}) is
rewritten in the "symmetric" form, setting $\sigma$ equal to
$\sigma_{xx}$ - the conductivity of a 2DEG in a magnetic field and
assuming the distance from the channel with the 2DEG to the
surface of the sample equals 0 (which corresponds to the
experimental structure), then $\Gamma$ in dB/cm is given by

\begin{eqnarray}
\Gamma=34.72b(k,a)(\varepsilon_1+\varepsilon_0)\varepsilon_0^2\varepsilon_s
e^{(-2ka)} \times \nonumber & \\ \times
K^2 k\frac { (\frac{4\pi \sigma_{xx}}
{\varepsilon_sv})r(k,a) } {1+[(\frac{4\pi
\sigma_{xx}}{\varepsilon_s v})r(k,a)]^2}, \label{Eq4}
\end{eqnarray}

where $\varepsilon$=51 (\cite{10}), $\varepsilon$=12, and
$\varepsilon_0$=1 are the permittivities of $LiNb0_3$, GaAs, and
of the vacuum gap between the sample and $LiNbO_3$ respectively;

$$b(k,a)={[c-t e^{(-2ka)}][c+m-t e^{( -2ka)} - p e{(
-2ka)}]}^{-1},$$ $$r(k,a)=\frac{c+m-t e^{(-2ka)}-p e^{(-2ka)}} {2[c-t
e^{(-2ka)}]},$$

where
$c=(\varepsilon_1+\varepsilon_0)(\varepsilon_s+\varepsilon_0)$,
$t=(\varepsilon_1-\varepsilon_0)(\varepsilon_s\varepsilon_0)$,
$m=(\varepsilon_1+\varepsilon_0)(\varepsilon_s-\varepsilon_0)$,
and $(\varepsilon_1-\varepsilon_0)(\varepsilon_s+\varepsilon_0)$.

Let us now analyze $\Gamma$ as a function of $\sigma_{xx}$ in
accordance with (\ref{Eq4}). If the conductivity is high, then

\begin{eqnarray}
(4\pi\sigma_{xx}/v\varepsilon_s)r(k,a)\gg 1
 \label{Eq5}
\end{eqnarray}

and $\Gamma\sim\sigma_{xx}$;if (for low conductivity)

\begin{eqnarray}
(4\pi\sigma_{xx}/v\varepsilon_s)r(k,a)\ll 1
 \label{Eq6}
\end{eqnarray}

then the absorption coefficient $\Gamma\sim\sigma_{xx}$.
Therefore, it is obvious that for

\begin{eqnarray}
(4\pi\sigma_{xx}/v\varepsilon_s)r(k,a)= 1
 \label{Eq7}
\end{eqnarray}

the function $\Gamma(H)$ possesses a maxim $\Gamma_M$, and from
(\ref{Eq4}) and (\ref{Eq7}) it is obvious that the value of
$\Gamma_M$ does not depend on the value of $\sigma_{xx}$. The
equality (7) in our case corresponds to

$$\sigma_{xx} r(k,a)=v$$

since $4\pi/\varepsilon_s\simeq 1$ for GaAs. Let us now consider
the dependence of the maximum absorption coefficient $\Gamma_M$ on
the wave frequency $f$, which follows from \ref{Eq4} and which is
shown in Fig. 3 for different values of $a$. It is obvious from
Fig. 3 that the dependence $\Gamma_M(f)$ has a maximum, and the
width of the vacuum gap increases with decreasing value of
$\Gamma_M$, so that the maximum occurs at a lower frequency.

If we now refer to Fig. 2, then on the basis of the fore-going
analysis of \ref{Eq4} it is possible to explain the splitting of
the maxima of $\Gamma$ at low filling numbers. Indeed, near the
conductivity maxima, where $\sigma_{xx}\simeq 10^{-5} \Omega^{-1}$
the condition  \ref{Eq5} is satisfied and the absorption
coefficient $\Gamma$ assumes its minimum values. As the magnetic
field increases, within the same Landau level the quantity
$\sigma_{xx}$ starts to decrease rapidly, and the coefficient
$\Gamma$ in accordance with \ref{Eq5} increases, until
$\sigma_{xx}$ decreases to a value corresponding to condition
(\ref{Eq7}). In this case $\Gamma(H)$ reaches its maximum value
$\Gamma_M$. As $\sigma_{xx}$ decreases further, the condition
(\ref{Eq6}) is satisfied and the coefficient $\Gamma$,
correspondingly, also decreases. This occurs right
 up to values of $H$ for which $\sigma_{xx} (H)$
 reaches its minimum value.
 As the magnetic field increases further, $\sigma_{xx}$
 again increases, the changes in $\Gamma$ occur in the reverse order.
 As a result, a second maximum appears in the dependence
 $\Gamma(Í)$.
 In weaker magnetic fields the quantity $\sigma_{xx}(H)$
 does not reach at the minima the values for which the equality (\ref{Eq7})
 holds. For this reason, splitting of the maxima of $\Gamma(Í)$
 does not occur.
 As the temperature decreases, the oscillations of
 $\sigma_{xx}(H)$
 become sharper and deeper,
 which is reflected in the corresponding curves for $\Gamma(H)$
 as a splitting of the maxima with the larger numbers.

To calculate from $\Gamma$ the dissipative conductivity
$\sigma_{xx}$, it is necessary to know the gap width a between the
insulator and the experimental object. In our case the
heterostructure is pressed directly to the lithium niobate surface
and a is a poorly controllable parameter, since it is determined
by irregularities of unknown amplitude on both surfaces. The
effective value of $a$ can be determined from acoustic
measurements by investigating the frequency dependence of the
maximum value of the absorption coefficient $\Gamma$. For the same
placement of the sample, i.e., for the same value of a but
different SAW frequencies, we obtain [see Eqs (\ref{Eq4}) and
(\ref{Eq7})]

\begin{eqnarray}
\frac{\Gamma(k_1)}{\Gamma(k_2)}=
  \frac{[k_1b(a,k_1)]}{[k_2 b(a,k_2)]}e^{-2a(k_1-k_2)}.
 \label{Eq8}
\end{eqnarray}

The value of $a$ determined from this equation was found to be
$0.25-1 \mu m$ for different placements of the sample. Knowing
$a$, it is possible to
 determine from relation (\ref{Eq4}) and the experimentally measured values
 of $\Gamma$ the conductivity $\sigma_{xx}^{ac}$ and
 its dependence on the magnetic field.
 It was found that this quantity, which was determined from the acoustic
  measurements for different values of a in the range of magnetic fields
  corresponding to absorption coefficient minima (conductivity maxima),
  corresponds within 20$\%$ to the values of $\sigma_{xx}^{dc}$
  (for the same values of $H$)
   which were calculated from the galvanomagnetic measurements of the
   dependences $\rho_{xx}(H)$ and $\rho_{xy}(H)$ in a direct current:

\begin{eqnarray}
\sigma_{xx}^{dc}(H)=\frac{\rho_{xx}(H)}{[\rho_{xy}(H)]^2+[\rho_{xx}(H)]^2}.
 \label{Eq9}
\end{eqnarray}

This equality apparently means (by analogy with the
three-dimensional case) that in the indicated range of magnetic
fields the static conductivity $\sigma_{xx}^{dc}$ is determined by
carriers in delocalized states.

To analyze further the experimental SAW absorption curves we
employed the vacuum gap $a$ as a parameter, choosing it so that
near the conductivity maxima $\sigma_{xx}^{ac}$ would equal
$\sigma_{xx}^{dc}$. In this method of determination, the vacuum
gap $a$ differs by no more than 20$\%$ from the gap calculated
from (\ref{Eq8}).

\section{DISCUSSION}

Figure 4 shows the magnetic field dependencies of the dissipative
conductivity $\sigma_{xx}^{ac}(H)$, which were obtained from the
SAW absorption coefficient for different frequencies and gap
widths $a$, as well as the dependence $\sigma_{xx}^{dc}(H)$ near
the maximum with filling number $\nu=nhc/2eH$=3.5, where $n$ is
 the Hall electron density. It should be noted that if the value of $a$
 determined by the method described in the preceding section
 is substituted into (\ref{Eq4}) and the value of $\sigma_{xx}^{dc}$
 measured in a direct
  current is used for $\sigma_{xx}$,
  then it is not possible to describe the entire
  experimental dependence $\Gamma(H)$.

It is obvious from Fig. 4 that the conductivities are equal to
each other in magnetic fields in the range 37-45 kOe, but as the
magnetic field moves away from the maximum in different
directions, they separate, and $\sigma_{xx}^{ac}$ is always
greater than $\sigma_{xx}^{dc}$. In our view, the difference of
the rf and static conductivities describes the transition through
the mobility threshold from free to localized electronic states
and correspondingly to different mechanisms of conduction in
static and rf electric fields.

To explain the experimental data on SAW absorption, we assume that
the Landau levels are smeared in space by the fluctuation
potential which may not be small. Indeed, in the best
GaAs/AlGaAs-type structures the mobility of the 2DEG at $T$=4.2K
reaches values of $\mu\simeq 10^7 cm^2/(V\cdot s)$, i.e., it is two
orders of magnitude greater than the mobility in the structure
investigated by us. The simplest explanation for the difference in
$\sigma_{xx}^{ac}$ and $\sigma_{xx}^{dc}$ could be as follows: Far
from the maxima the conductivity $\sigma_{xx}$ is of the character
of hops along localized states and should therefore increase with
frequency $f$ as $\sigma \sim f^s (s=1)$ \cite{1}.  However, as we
shall show below, this is at variance with the experimental
conclusion that in measurements of $\Gamma$ at different
frequencies and calculations of $\sigma_{xx}^{ac}$ from them we
did not observe a frequency dependence of $\sigma_{xx}^{ac}$
within the limits of the experimental error, i.e., according to
our data $\sigma =0$. Therefore, a more adequate model is the
model of a large-scale fluctuation potential and the assumption
that for 2DEG a "mobility threshold" exists in our samples. As a
result, in the range of magnetic fields where the Fermi level lies
above the mobility threshold (region of maximum $\sigma_{xx}$),
the electrons are delocalized and
$\sigma_{xx}^{ac}(H)=\sigma_{xx}^{dc}(H)$. As the magnetic field
varies, the Fermi level crosses the mobility threshold, causing a
percolation-type transition, in which the electrons are localized
in the random potential, forming so-called "lakes" with
metallic-type conductivity. In the latter case the static
conductivity mechanism becomes different from the rf conductivity
mechanism: The static conductivity is determined by the activation
transfer of electrons to the percolation level and the rf
conductivity, whose magnitude is larger than that of the static
conductivity, is determined by the conductivity in the lakes
(along closed orbits for free 2D electrons, determined by the
relief formed by fluctuations of the potential energy). This model
is supported by the fact that in the range of magnetic fields
where $\sigma_{xx}^{ac}(H) = \sigma_{xx}^{dc}(H)$ the quantity
$\rho_{xy}$ increases abruptly at a transition between two quantum
plateaus (Fig. 4), which corresponds to the position of the Fermi
level in the region of delocalized states \cite{11}.

It is very difficult to calculate the SAW absorption coefficient
of 2D electrons localized in a random potential. For this reason,
we assume, by analogy with the three-dimensional case \cite{3}
that for the two-dimensional conductivity the formula for
$\Gamma$, which describes SAW absorption in this case, will have
the same structure as (\ref{Eq4}), but a factor $\Sigma$, which
depends on the magnetic field and temperature, appears in this
case in front of the fraction. For a three-dimensional electron
gas, the physical meaning of the coefficient $\Sigma(H)$ is that
of a relative volume occupied by the conducting "drops," whereas
in our case (2DEG) $\Sigma(H)$ is approximately proportional to
the density of electrons which are localized on the minima of the
fluctuation potential (lakes) and which realize in the lakes a
conduction in the electric field of the SAW. As a result,

\begin{eqnarray}
\Gamma= \Sigma(H)\Gamma(\sigma_{xx}^l),
 \label{Eq10}
\end{eqnarray}

where $\Gamma(\sigma_{xx}^l)$ corresponds to expression
(\ref{Eq4}) with $\sigma_{xx}$ replaced by $\sigma_{xx}^{l}$—the
conductivity in the lakes. In the range of fields $H$ where the
electrons are delocalized, the function $\Sigma(H)$ = 1 and

$$\sigma_{xx}^{ac}(H)=\sigma_{xx}^{ac}(H),$$

and then the value of $\Sigma$ becomes less than 1. Therefore, to
calculate $\sigma_{xx}^{ac}(H)$ in the entire range of magnetic
fields, it is necessary to know the value of $\Sigma$ and its
magnetic-field dependence.

The following operation was performed to determine $\Sigma(H)$:
The ratio of the experimentally measured values of $\Gamma(H)$ and
$\Gamma(H)$ was determined near each maximum of $\sigma_{xx}$ with
respect to the magnetic field in the range of magnetic fields
where $\Gamma \sim l/\sigma$, assuming that near the percolation
level $\sigma_{xx}^{ac}=\sigma_{xx}^{ac}$. Then

\begin{eqnarray}
\frac{\Gamma(H)}{\Gamma(H_{max})}=
  \frac{\Sigma(H)\sigma_{xx}^{dc}(H_{max})}{\sigma_{xx}^{dc}(H)}.
 \label{Eq11}
\end{eqnarray}

since $\Gamma(H)$ is determined according to (\ref{Eq4}). As
indicated above, the ratio
$\sigma_{xx}^{dc}(H_{max})/\sigma_{xx}^{dc}(H)$ is calculated from
the components $\rho_{xx}(H)$ and $\rho_{xy}(H)$ of the
resistivity tensor which are measured for this sample.

Another method to obtain $\Sigma$ was to determine it from the
quantity $\Gamma_M$ (see Fig. 2). This method did not require a
knowledge of $\sigma_{xx}^{ac}$, since we obtain from Eqs. (4),
(7), and (10) the expression

\begin{eqnarray}
\Gamma=\Sigma(H) 17.36 K^2k exp(-2ka)(\varepsilon_1+\varepsilon_0)
\varepsilon_s\varepsilon_0^2b(k,a).
 \label{Eq12}
\end{eqnarray}

The dependence of the function $\Sigma$, which is determined by
the methods described above, on $|\Delta\nu|$ at $T$=4.2 Ê in the
magnetic field range corresponding to filling numbers $\nu$ from 3
to 4, where $\Delta\nu$ is measured from $\nu$=3.5, is shown in
Fig. 5. It should be noted that the points on the plot were
obtained from the values of the absorption coefficient for
different SAW frequencies and different values of $a$. The value
of $\Sigma$ depends on the temperature and is a subject of our
further investigations. The solid line in Fig. 5 was estimated by
eye and extrapolated to the region $\Delta\nu$= 0.5, so that it
was impossible to calculate the value of $\Sigma$ near the
conductivity minima because of the lack of accurate data on the
value of the static conductivity in the limit
$\sigma_{xx}^{dc}\rightarrow 0$.

Knowing the vacuum gap and the function $\Sigma(H)$, the value of
$\Sigma(H)$ could be determined according to Eqs. (4) and (10)
from the experimental values of $\Gamma$. The results of such
calculations are presented in Fig. 6 for $T$=4.2Ê. The different
points correspond to curves obtained at different frequencies and
with different vacuum gaps. The solid line represents the function
$\sigma_{xx}^{dc}(H)$. As one can see from Fig. 6, for large
Landau numbers $\sigma_{xx}^{ac}(H)=\sigma_{xx}^{dc}(H)$, just as
in the weaker magnetic fields, but only near the conductivity
maxima. In the regions of the magnetic field where
$\sigma_{xx}^{dc}\rightarrow 0$, however, the values of
$\sigma_{xx}^{dc}$ and $\sigma_{xx}^{ac}$ diverge, and
$\sigma_{xx}^{ac}$ is always greater than $\sigma_{xx}^{dc}$, as
should be the case on the basis of the model proposed above. The
equality $\sigma_{xx}^{ac}(H)=\sigma_{xx}^{dc}(H)$ for high
Landau-level numbers is attributable to the fact that the mobility
thresholds at $T$=4.2K for neighboring levels are smeared in
energy and overlap, so that the condition for carrier localization
is not achieved in weak magnetic fields, even when the chemical
potential falls between the Landau levels.

Since $\sigma_{xx}^{ac}$ is calculated from $\Gamma$ after a
rather complicated procedure of analysis of the experimental data,
we estimate the error in $\sigma_{xx}^{ac}$ to be of the order of
40$\%$. To this accuracy, the quantity $\sigma_{xx}^{ac}$ at the
minima with respect to the magnetic field for small filling
numbers, i.e., in localized states, does not depend on the sound
frequency (and in our experiments the frequency varied by a factor
of 7).

It should be noted that our results are at variance with the data
of \cite{12}, where $\sigma_{xx}^{ac}$ was studied in the Corbino
disk geometry in the frequency range 50-600 MHz. The authors found
that $\sigma_{xx}^{ac}$ does not equal to $\sigma_{xx}^{dc}$
either at the maxima or at the minima of the oscillations of
$\sigma_{xx}(H)$ when the filling numbers $\nu$ are small. They
explained the observed frequency dependence of $\sigma_{xx}$ in
terms of the theory given in \cite{13}.

\section{CONCLUSIONS}

The magnetic field and frequency dependencies of the
high-frequency conductivity of the 2DEG in the heterostructure
GaAs/AlGaAs (in the region 30-210 MHz) were calculated and
analyzed. It was shown that the experimental results can be
explained if it is assumed that there exists a fluctuation
potential in which carrier localization occurs. The character of
the SAW absorption in the presence of an interaction with
localized carriers is discussed.

\section{ACKNOWLEDGEMENTS}

We thank Yu. M. Gal'perin, V. D. Kagan, and A. Ya. Shik for
helpful discussions and G. O. Andrianov for assisting in the work.

This work was supported by the Russian Fund for Fundamental
Research Grants Nos. 95-02-0466a and 95-02-04042a as well as INTAS
Grants Nos. 93-1403, 93-1403-EXT, and 95-IN/RU-553.

\end{multicols}
%\widetext

\newpage

\begin{figure}[t]
\centerline{\psfig{figure=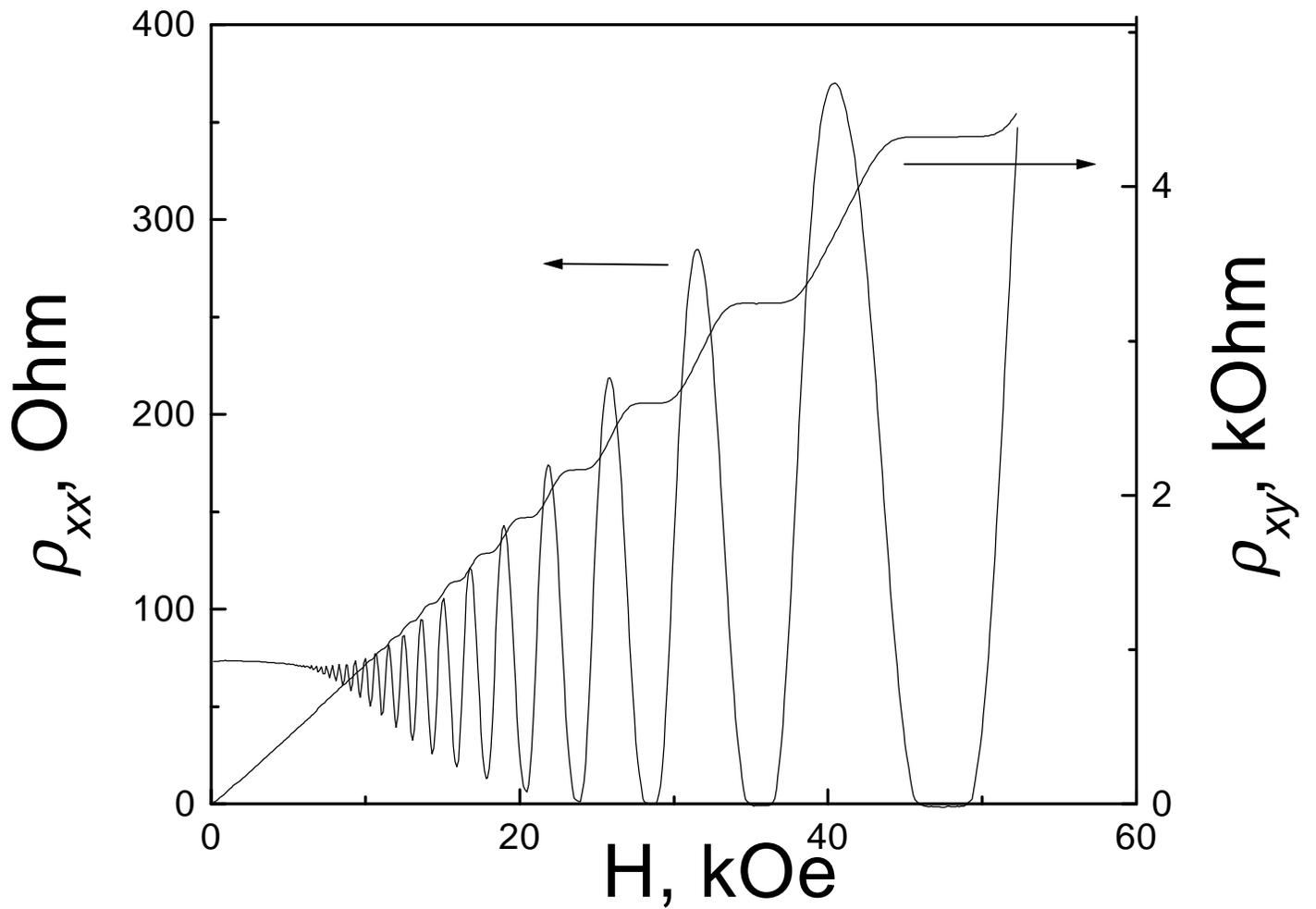}} \caption{ Resistivity
tensor components $\rho_{xx}$ and $\rho_{xy}$ versus the magnetic
field $H$ in a GaAs/AlGaAs heterostructure at $T$=1.5K.
\label{fig1}}
\end{figure}

\newpage

\begin{figure}[h]
\centerline{\psfig{figure=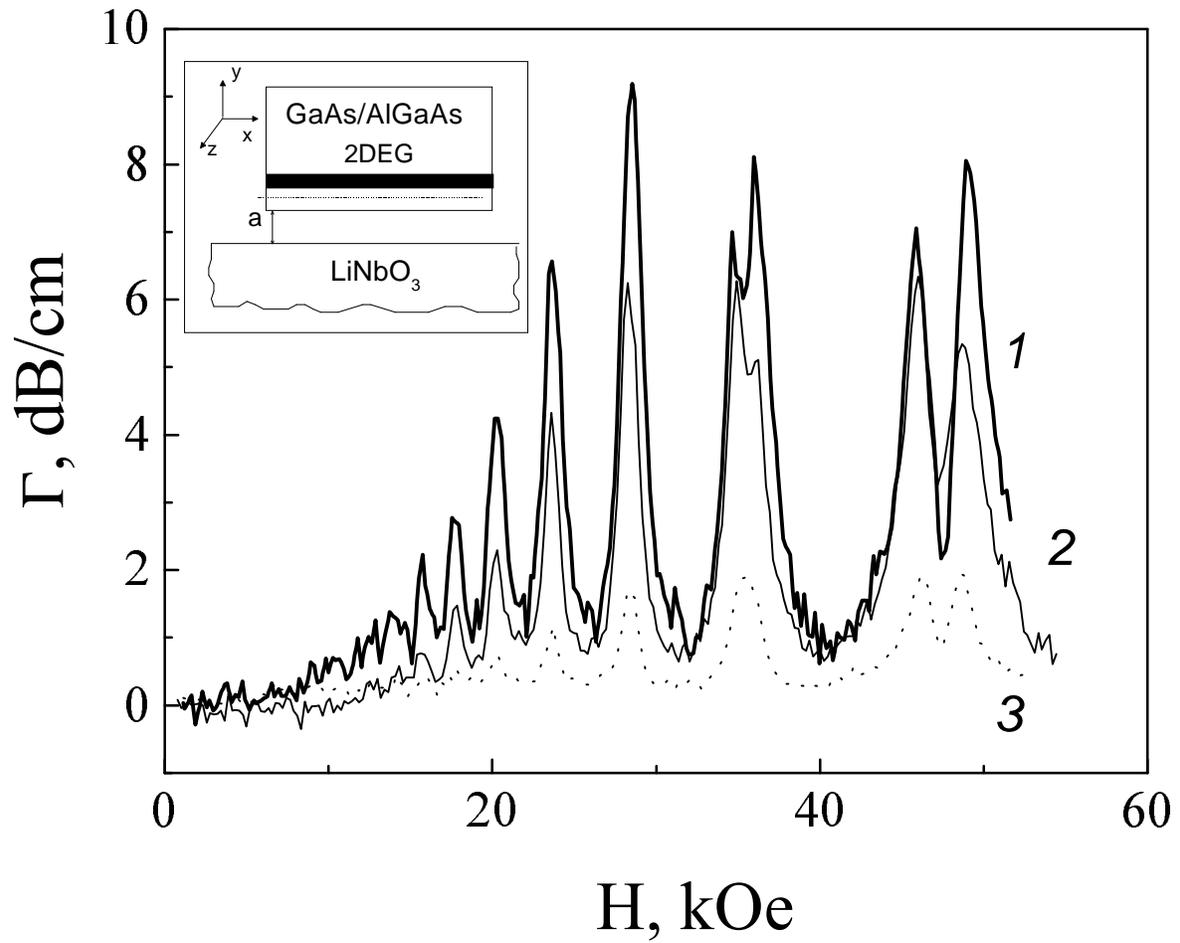}} \caption{Surface acoustic
wave absorption coefficient $\Gamma$ versus the magnetic field
intensity $H$ at $f$=30MHz with vacuum gap widths $a$, $\mu m$: 1
- 0.3, 2 - 0.5, and 3 - 1.0. Inset: Arrangement of the sample with
respect to the crystal axes of lithium niobate. \label{fig2}}
\end{figure}

\newpage

\begin{figure}[h]
\centerline{\psfig{figure=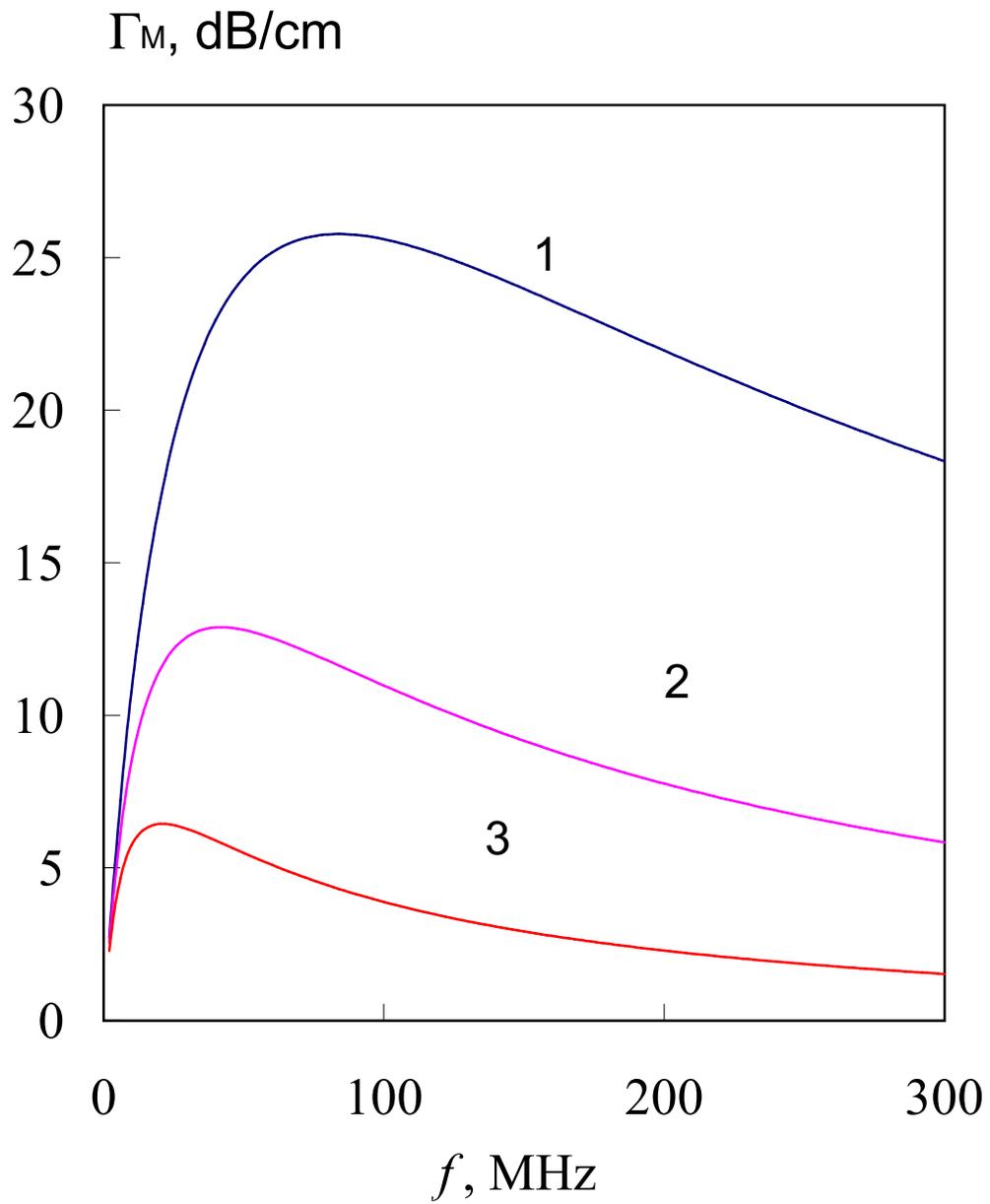}} \caption{Theoretical value
of the maximum absorption $\Gamma$ versus acoustic wave frequency
$f$ for different vacuum gap widths $a$, $\mu m$: 1 - 0.3, 2 - 0.6,
3 - 1.2. \label{fig3}}
\end{figure}

\newpage

\begin{figure}[h]
\centerline{\psfig{figure=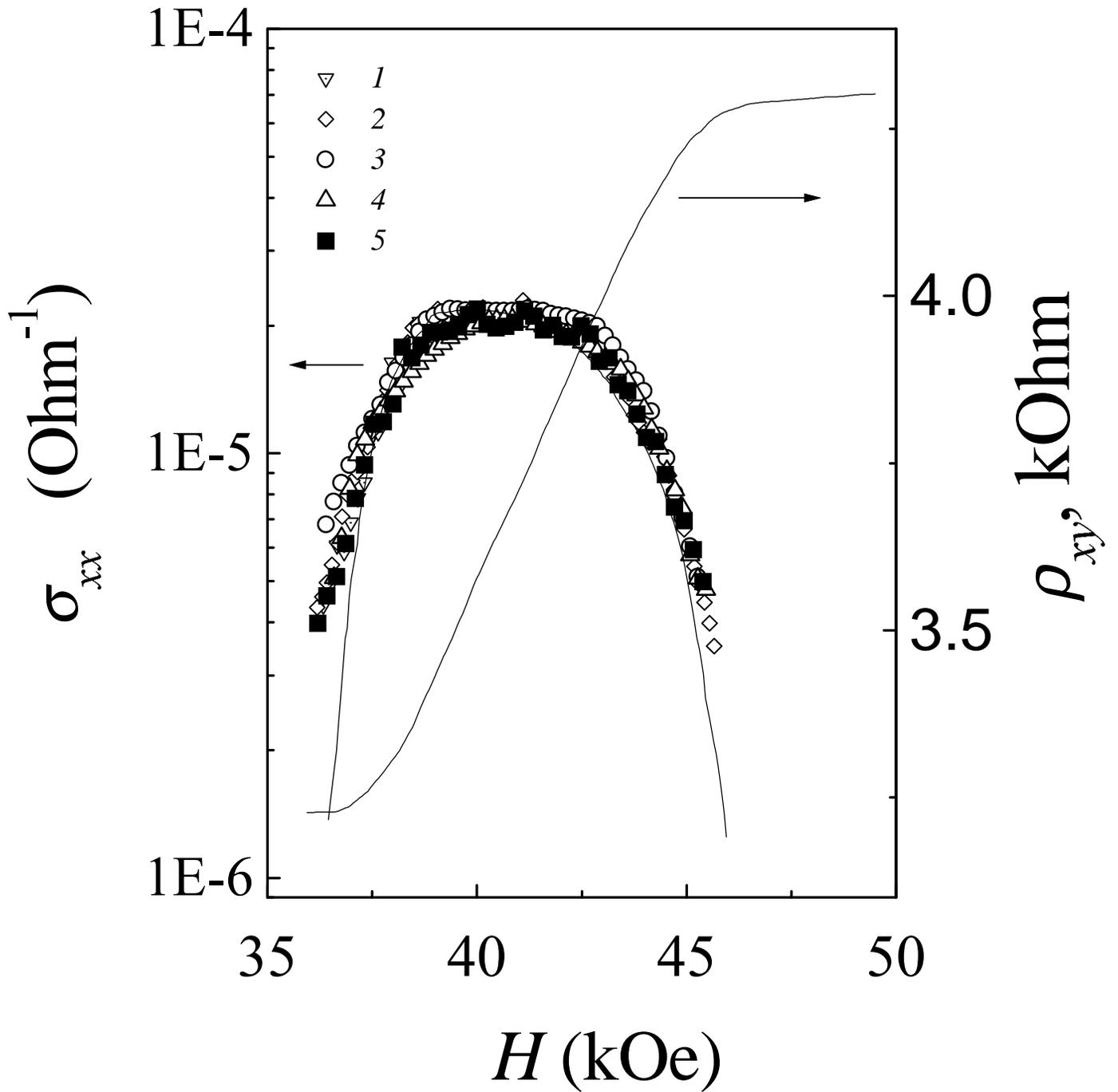}}
\caption{$\sigma_{xx}^{dc}$ (solid line) and
$\sigma_{xx}^{ac}$ (symbols) (1) versus the magnetic
field $H$ near the region of delocalized states: $\rho_{xy}(H)$.
The symbols correspond to frequencies $f$ (MHz) and vacuum gap
widths $a$ (in $\mu m$): 3 - 213 and 0.3, 4 - 30 and 0.5, 5 - 150
and 0.3, 6 - 30 and 0.4, 7 - 90 and 1.2.\label{fig4}}
\end{figure}

\begin{figure}[h]
\centerline{\psfig{figure=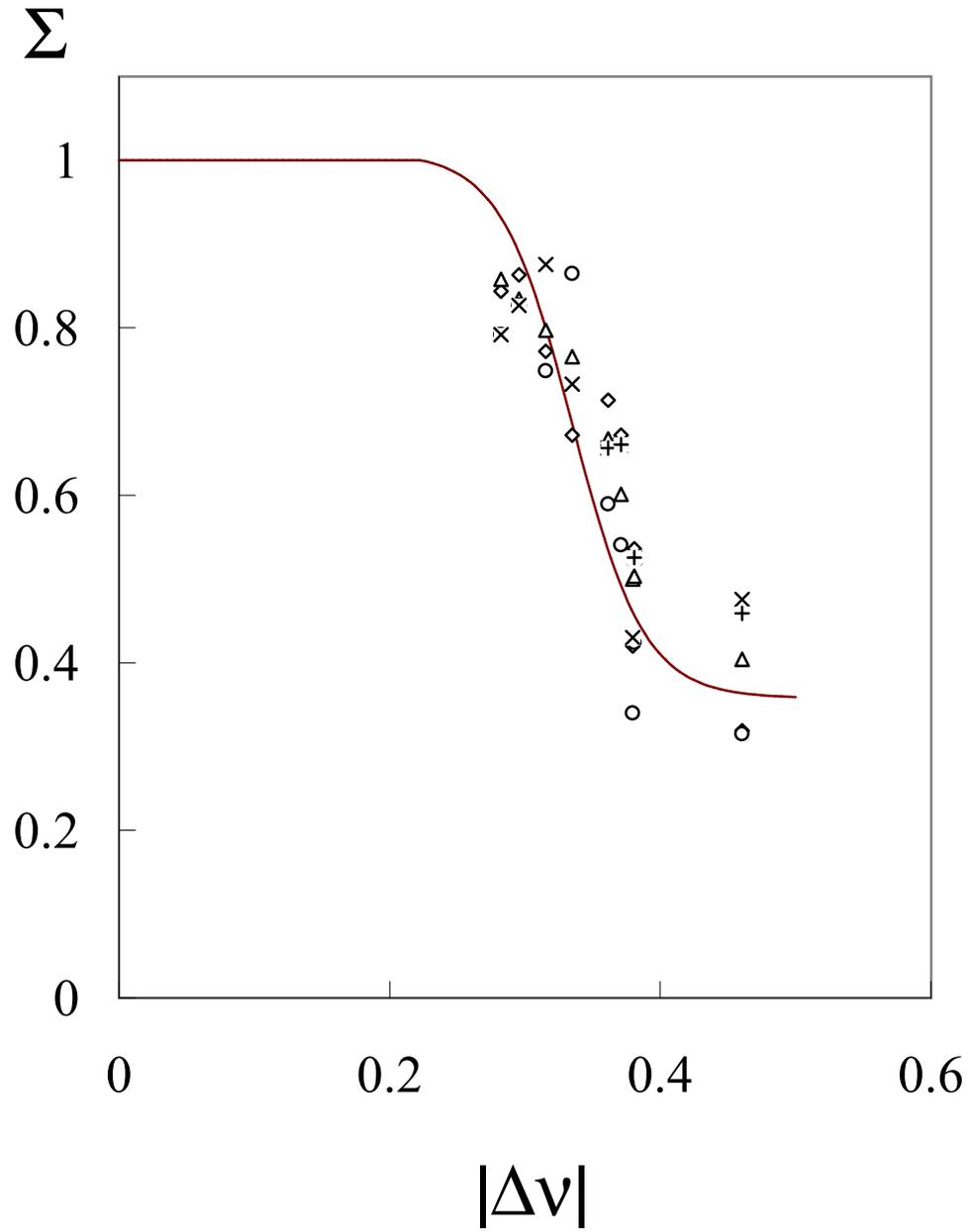}} \caption{$\Sigma$ versus
$|\Delta\nu|$ - absolute value of the change in the filling factor
$\nu$, measured from $\nu$=3.5. The symbols correspond to
frequencies $f$ (MHz) and vacuum gap widths $a$ (in $\mu m$): 1 -
213 and 0.3, 2 - 30 and 0.5, 3 - 50 and 0.3, 4 - 30 and 0.4, 5 -
90 and 1.2.\label{fig5}}
\end{figure}
\newpage

\begin{figure}[h]
\centerline{\psfig{figure=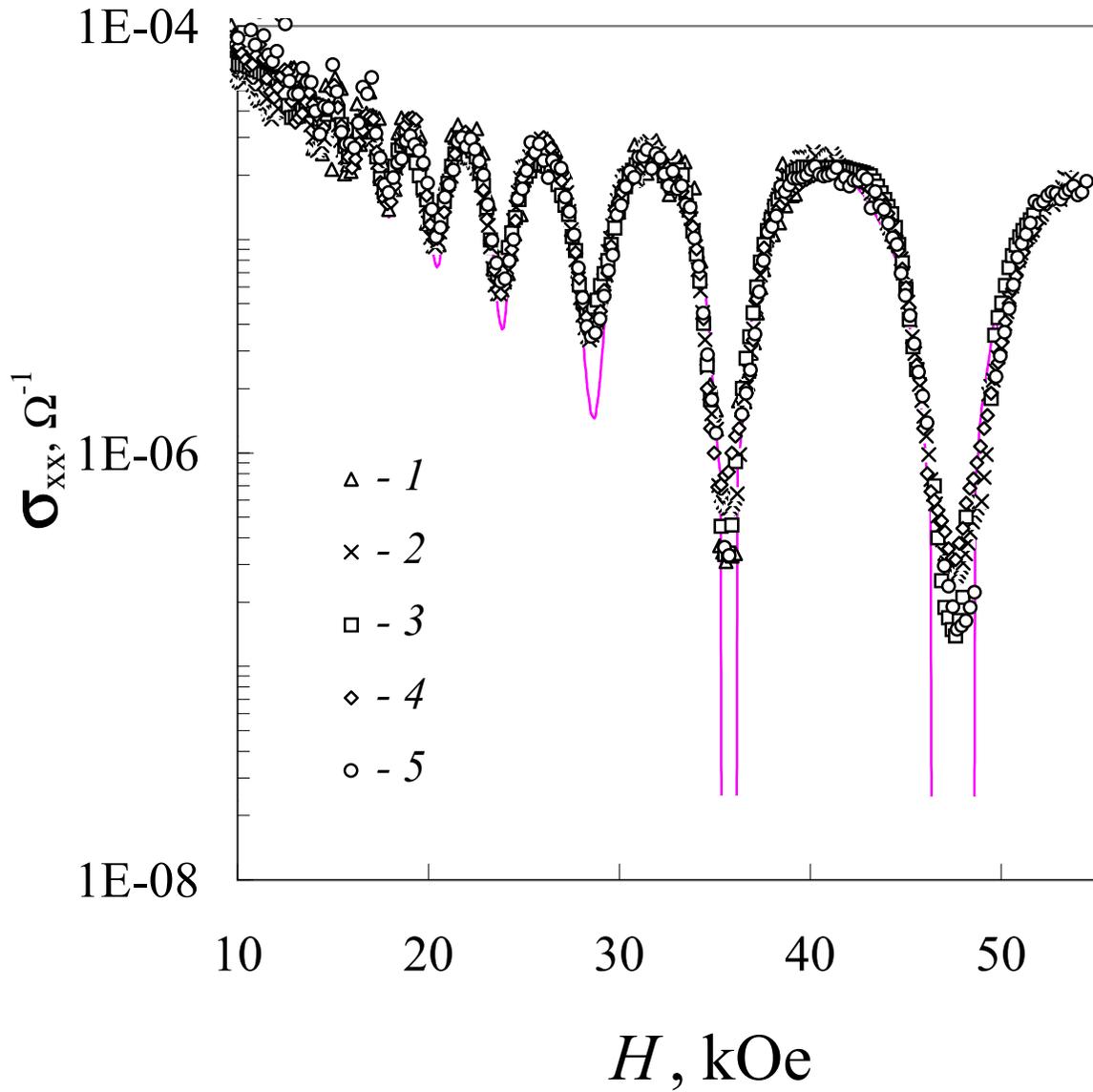}} \caption{
$\sigma_{xx}^{dc}$ (solid line) and $\sigma_{xx}^{ac}$ (symbols)
versus the magnetic field $H$. The different symbols have the same
meaning as in Fig. 5. \label{fig6}}
\end{figure}

\end{document}